\documentclass[12pt]{iopart}

\usepackage{iopams}  

\usepackage{graphicx}
\usepackage{float}
\usepackage{bm}
\usepackage{bbold}
\usepackage[caption=false]{subfig}
\usepackage{hyperref}
\hypersetup{colorlinks,%
citecolor=blue,%
filecolor=black,%
linkcolor=blue,%
urlcolor=blue
}

\usepackage{dsfont}

\newcommand{\bra}[1]{ \langle #1 \vert}
\newcommand{\ket}[1]{ \vert #1 \rangle}
\newcommand{\kb}[2]{\vert #1 \rangle \langle #2 \vert}
\newcommand{\eqref}[1]{(\ref(#1))}
\newcommand{\citep}[1]{\cite(#1)}
\newcommand{\floor}[1]{\left\lfloor #1 \right\rfloor}

\begin{document}

\title[ML approach to reconstruct Density matrices from Quantum marginals]{Machine Learning approach to reconstruct Density matrices from Quantum Marginals}
\author{Daniel Uzcategui-Contreras}
\ead{daniel.uzcategui@ua.cl}
\address{
  Departamento de Física, Facultad de Ciencias Físicas y Matemáticas, Universidad de Concepción, Concepción, Chile \\
  Millennium Institute for Research in Optics (MIRO), Chile
}
\author{Antonio Guerra}
\address{
  Departamento de Física, Facultad de Ciencias Físicas y Matemáticas, Universidad de Concepción, Concepción, Chile \\
  Millennium Institute for Research in Optics (MIRO), Chile
}
\author{Sebastian Niklitschek}
\address{
  Departamento de Estadítica, Facultad de Ciencias Físicas y Matemáticas,
  Universidad de Concepción, Concepción, Chile
}
\author{Aldo Delgado}
\address{
  Departamento de Física, Facultad de Ciencias Físicas y Matemáticas, Universidad de Concepción, Concepción, Chile \\
  Millennium Institute for Research in Optics (MIRO), Chile
}

\date{\today}
\begin{abstract}
In this work, we propose a machine learning-based approach to address a specific aspect of the Quantum Marginal Problem: reconstructing a global density matrix compatible with a given set of quantum marginals. Our method integrates a quantum marginal imposition technique with convolutional denoising autoencoders. The loss function is carefully designed to enforce essential physical constraints, including Hermiticity, positivity, and normalization. Through extensive numerical simulations, we demonstrate the effectiveness of our approach, achieving high success rates and accuracy. Furthermore, we show that, in many cases, our model offers a faster alternative to state-of-the-art semidefinite programming solvers without compromising solution quality. These results highlight the potential of machine learning techniques for solving complex problems in quantum mechanics.
\end{abstract}

\section{Introduction}
\label{sec:intro}
Machine Learning (ML) has demonstrated remarkable capabilities across a broad range of scientific and technological fields, from classical tasks like regression and classification to sophisticated applications such as large language models (LLMs). Its success in discovering complex, nonlinear relationships has accelerated advances in fundamental science \cite{karagiorgi_2022,PhysRevResearch.2.033429}, mathematics \cite{ellenberg_2024}, materials science \cite{deep_mind_materials_2023}, and meteorology \cite{doi:10.1126/science.adi2336}. In quantum mechanics, ML techniques have been applied to tasks such as quantum state tomography \cite{schmale_qst_2022,carrasquilla_qst_2019} and quantum process tomography \cite{giuseppe_qpt_2023}, both of which suffer from exponential scaling with system size, rendering them impractical for large systems. Additionally, physics-informed neural networks (PINNs) \cite{raissi_pin_2019} have shown promise for solving quantum control problems through the resolution of partial differential equations like the Schrödinger equation \cite{norambuena_2024}.

In this context, the Quantum Marginal Problem (QMP) arises as a fundamental challenge in quantum theory: Given a set of reduced density matrices describing subsystems of a multipartite system, under what conditions are they consistent? Here, consistent means that there is a global density matrix, describing the entire system, that contains the reduced density matrices. This question naturally leads to several others: Can such a global state be uniquely determined? How is entanglement reflected in the marginals? What is the minimal information required to reconstruct the global state? 

The consistency problem of the QMP is a decision problem that was shown to be complete for the Quantum Merlin-Arthur (QMA) complexity class \cite{10.1007/11830924_40,doi:10.1137/21M140729X}, highlighting its computational intractability in the general case. Solving the QMP would have profound implications. For example, it could enable more efficient calculations of the ground states of local Hamiltonians \cite{PhysRevLett.98.110503}. The QMP also plays a central role in quantum chemistry, where it is referred to as the N-representability problem. Understanding the conditions under which reduced wave functions of electron pairs are compatible with an N-electron system could lead to more efficient calculations of binding energies \cite{10.1063/1.1359715}. However, the N-representability problem is also known to be QMA complete, reflecting its computational difficulty \cite{PhysRevLett.98.110503}.

Previous works have addressed special cases of the QMP. For example, Klyachko provided the necessary and sufficient conditions for compatibility of 1-body marginals with a global pure state in the nonoverlapping case \cite{Klyachko2004}, while other studies have explored the uniqueness of pure global states determined by reduced marginals \cite{wootters2002,Lajos2004,FelixNikolai2017}. Additionally, symmetric instances of the QMP have been formulated as semidefinite programming (SDP) problems \cite{aloy2020quantum}.

Machine learning (ML) approaches have been proposed for problems in quantum many-body physics \cite{carleo_qmbp_2016,carleo_mlp_2019,preskil_ml_2022,PhysRevResearch.5.013097}. Deep learning models have been shown to learn $N$-representability conditions from one- and two-electron systems and predict properties of larger systems \cite{doi:10.1021/jacs.2c07112,doi:10.1021/acs.jpclett.4c03366,shao_2023}. In this work, we introduce a novel deep learning-based approach to approximate solutions to QMP. Our method combines the Marginal Imposition Operator (MIO) \cite{uzcateguic_2022}, a technique designed to impose arbitrary quantum marginals on a matrix, with a convolutional denoising autoencoder (CDAE) architecture. This combination enables the construction of physically valid quantum states (i.e., positive semidefinite with trace one) that are compatible with a prescribed set of marginals, even in the presence of noise or initial inconsistency.

Our architecture is capable of handling mixed quantum states for systems of 3 to 8 qubits. Additionally, we explore transfer learning between models trained on different numbers of qubits, showing that a model trained on 3-qubit systems can be adapted to systems with up to 8 qubits with limited retraining. This indicates that the network captures structural features of the compatibility conditions that are not specific to a fixed Hilbert space dimension.

The results presented in this article suggest that deep neural networks are powerful tools for approximating feasible solutions to the QMP, and they may offer a scalable alternative to traditional optimization-based methods. Furthermore, by integrating explainability techniques such as knowledge extraction \cite{boger_1997,HORTA2021109,odense2020layerwiseknowledgeextractiondeep,olah2020zoom,olah2017feature}, we believe that these models can contribute to a deeper theoretical understanding of the quantum marginal landscape.

\section{The problem of compatibility}
\label{sec:compatibility_problem}
A quantum marginal $\sigma_{\mathcal{J}}$ is the reduced quantum state of a subsystem labeled by $\mathcal{J}$, derived from a larger system with global state $\rho_{\mathcal{I}}$. Let $\mathcal{I}$ denote a set of labels (e.g., $\mathcal{I} = {1, \ldots, N}$) representing the components of an $N$-body physical system. The subset $\mathcal{J} \subset \mathcal{I}$ corresponds to the components of a reduced subsystem. When the global state $\rho_{\mathcal{I}}$ is known, the marginal state $\sigma_{\mathcal{J}}$ can be obtained via the partial trace operation as:
\begin{equation}
\sigma_{\mathcal{J}} = \mathrm{tr}_{\mathcal{J}^{c}} \left[ \rho_{\mathcal{I}} \right] = \sum_j \left(\bra{j^c} \otimes\mathbb{1}_{\mathcal{J}}\right)\rho_{\mathcal{I}} \left( \ket{j^c} \otimes\mathbb{1}_{\mathcal{J}} \right),
\label{patrace}
\end{equation}
where $\mathcal{J} \cup \mathcal{J}^c = \mathcal{I}$. In equation (\ref{patrace}), the identity operator $\mathbb{1}_{\mathcal{J}}$ acts on the subsystem $\mathcal{J}$, while $\{ \ket{j^c} \}$ is an orthonormal basis tracing the complement $\mathcal{J}^c$. 

Let us consider now the opposite problem. This is, given a set of $M$ quantum marginals $\{ \sigma_{\mathcal{J}_{\alpha}} \}_{\alpha=1}^M$, we would like to know whether there is a global quantum state $\rho_{\mathcal{I}}$  such that $\sigma_{\mathcal{J}_{\alpha}} = \mathrm{tr}_{\mathcal{J}^{c}_{\alpha}} \left[ \rho_{\mathcal{I}} \right]$ for all $\alpha=1,\ldots,M$. This problem can be stated as the SDP \thinspace\cite{cavalcanti_sdp_2023}
\begin{eqnarray}
  \label{SDP-primal}
  \mathrm{maximize} &\quad& \mathrm{tr}(\rho_{\mathcal{I}}) \nonumber \\
  \mathrm{subject\ to} &\quad& \mathrm{tr}_{\mathcal{J}^{c}_{\alpha}} \left[ \rho_{\mathcal{I}} \right] = \sigma_{\mathcal{J}_{\alpha}}, \quad \alpha = 1,\ldots,M \\
  &\quad& \rho_{\mathcal{I}} \geq 0 \nonumber
\end{eqnarray}
There are algorithms, such as interior point methods, that are highly efficient for solving SDPs. However, the runtime and memory requirements of these methods scale with the size of the matrix and the number of constraints \cite{jiang_sdpcomplexity_2020,zhang_2018}, making them impractical for large-scale instances of the problem. An iterative algorithm based on the Marginal MIO, which will be introduced in the next subsection, was proposed in \cite{uzcateguic_2022}. Nevertheless, determining the computational complexity of this approach remains a challenging task.

\subsection{The Marginal Imposition Operator}
The problem of determining a global quantum state from a given set of quantum marginals has been previously studied \cite{uzcateguic_2022}, where the main tool introduced is the MIO. Let $\sigma_{\mathcal{J}}$ be a $|\mathcal{J}|$-party quantum state, with $\mathcal{J}\subset\mathcal{I}$. Then, the MIO is defined as follows:
\begin{equation}\label{mio}
\mathcal{Q}_{\mathcal{J}}(\tilde{\rho}_{\mathcal{I}}):=\tilde{\rho}_{\mathcal{I}}-\rho_{\mathcal{J}} \otimes \frac{\mathbb{1}_{\mathcal{J}^{c}}}{|\mathcal{J}^c|}+\sigma_{\mathcal{J}} \otimes \frac{\mathbb{1}_{\mathcal{J}^{c}}}{|\mathcal{J}^c|},
\end{equation}
\noindent where $\tilde{\rho}_{\mathcal{I}}$ is an $|\mathcal{I}|$-party quantum state, which can be chosen at random or in any convenient fashion, and $\rho_{\mathcal{J}} = \mathrm{tr}_{\mathcal{J}^{c}} \left[ \tilde{\rho}_{\mathcal{I}} \right]$.

It is easy to see that $\mathrm{tr}_{\mathcal{J}^{c}} \left[ \mathcal{Q}_{\mathcal{J}}(\tilde{\rho}_{\mathcal{I}}) \right] = \sigma_{\mathcal{J}}$. In simple terms, the MIO enforces the marginal $\sigma_{\mathcal{J}}$ onto $\tilde{\rho}_{\mathcal{I}}$, although $\mathcal{Q}_{\mathcal{J}}(\tilde{\rho}_{\mathcal{I}})$ may not necessarily represent a valid quantum state. Despite this, the MIO exhibits several interesting properties:
\begin{enumerate}
    \item Hermitian: $\mathcal{Q}_{\mathcal{J}}(\tilde{\rho}_{\mathcal{I}}) = \mathcal{Q}_{\mathcal{J}}(\tilde{\rho}_{\mathcal{I}})^{\dagger}$
    \item  Trace preserving: $\mathrm{tr}\left[ \mathcal{Q}_{\mathcal{J}}(\tilde{\rho}_{\mathcal{I}}) \right]=1$.
\end{enumerate}

More generally, for a given set $\{ \sigma_{\mathcal{J}_{\alpha}} \}_{\alpha=1}^M$, the composition $\mathcal{Q}_{\mathcal{J}_M}\circ\dots\circ\mathcal{Q}_{\mathcal{J}_1}( \tilde{\rho}_{\mathcal{I}} )$ contains all the quantum marginals. This is 
\begin{equation}\label{mio_composition}
    \sigma_{\mathcal{J}_{\alpha}} = \mathrm{tr}_{\mathcal{J}^{c}_{\alpha}} \left[ \mathcal{Q}_{\mathcal{J}_M}\circ\dots\circ\mathcal{Q}_{\mathcal{J}_1}( \tilde{\rho}_{\mathcal{I}} ) \right],
\end{equation}
for all $\quad \alpha=1,\ldots,M$. An obvious condition for equation (\ref{mio_composition}) to be satisfied is that for every pair $\sigma_{\mathcal{J}_{\beta}}, \sigma_{\mathcal{J}_{\nu}} \in \{ \sigma_{\mathcal{J}_{\alpha}} \}_{\alpha=1}^M$, the quantum states $\sigma_{ \mathcal{J}_{\beta} \cap \mathcal{J}_{\nu} }$ in the overlapping subsystems must coincide when $\mathcal{J}_{\beta} \cap \mathcal{J}_{\nu} \neq \emptyset$.

However, there are no guarantees that $\mathcal{Q}_{\mathcal{J}M}\circ\dots\circ\mathcal{Q}_{\mathcal{J}1}( \tilde{\rho}_{\mathcal{I}} ) \geq 0$. Ensuring positivity may depend on the choice of $\tilde{\rho}_{\mathcal{I}}$, as numerically demonstrated in \cite{uzcateguic_2022}, among other factors. Although one could attempt to find the closest quantum state \cite{guta_2020}, this approach incurs the computational cost of eigendecomposition. In the next section, we introduce CDAE, an ML model that offers a potential solution to this problem.

\section{Model description}
\subsection{Image processing}
\label{sec:image-processing}
ML has emerged as a powerful tool for image processing \cite{image4}, a field encompassing techniques designed to transform an image into another representation based on a specific objective. The resulting output can be either another image or a classification/discrimination result. Examples of such transformations include resolution enhancement, noise reduction \cite{image2}, image segmentation \cite{image1}, and other feature extraction methods. A central goal of image processing is to preserve essential information while enhancing the relevant features of the original image.

A color image can be represented as a set of three matrices corresponding to the RGB color channels, each with the same dimensions, containing a fixed number of pixels in both height and width. This establishes a direct and intuitive correspondence between images and matrices.

Convolutional Neural Networks (CNNs) are designed to process multi-array-type data. For example, color images, which are represented as a set of three matrices corresponding to the RGB color channels, each with the same dimensions, containing a fixed number of pixels in both height and width. Many state-of-the-art models are built upon CNN architectures, including Autoencoders, U-Nets, Variational Autoencoders (VAEs), and SegNet, each tailored for specific tasks. These models are widely used in applications such as denoising, segmentation, and feature extraction. A key property of CNNs is their ability to progressively capture hierarchical features while reducing the dimensionality of input images, enabling efficient and effective processing.

On the other hand, a fundamental concept in quantum mechanics is the density matrix, which encodes the information of a quantum system. These matrices consist of complex-valued elements and must satisfy specific mathematical properties: they must be PSD and have a unit trace. Consequently, they are Hermitian. For a quantum system formed by $N$ qubits, the dimension of the space of density matrices grows exponentially with $N$.

A density matrix can be decomposed into its real and imaginary components, forming a multiarray: a set of two real-valued matrices. This suggests a natural analogy between density matrices and images, which in turn implies that image processing techniques may be adapted to quantum data. If CNNs are effective at extracting spatial features from images, it is reasonable to explore their application in learning quantum states.

However, this relationship is not straightforward. The convolutional layers of CNNs exhibit local connectivity, where each neuron is linked only to its neighboring units. This contrasts with fully connected layers, in which each neuron in one layer is connected to all units in the next layer. This localized connectivity is particularly beneficial in image processing, as it allows the extraction of relevant features while preserving spatial relationships, without requiring a global representation of the entire image.

In contrast, quantum density matrices do not inherently exhibit this locality-based structure. However, in QMP, we are primarily interested in global quantum states and their marginals, which are obtained through partial trace operations. This approach bears a conceptual similarity to the locality-based feature extraction performed in CNNs: While CNNs capture local structures within an image, partial traces extract relevant information from subsystems of a quantum state while maintaining a connection to the global system. This suggests that a CNN-based approach may be suitable for learning quantum states.

One of the main advantages of Deep Learning models for image processing is their scalability, as they are designed to handle images containing thousands or even millions of pixels. This scalability is particularly relevant in quantum mechanics, where the Hilbert space grows exponentially with the number of qubits. In this work, we present results for systems between 3 and 8 qubits. Training models for larger quantum systems requires significantly higher computational resources, which are currently beyond our reach. Nevertheless, we provide all the necessary tools for the further development of such models, facilitating their application to more complex quantum systems.

\subsection{Autoencoders}
\label{sec:autoencoders}
Autoencoders are neural networks designed to learn efficient representations of data \cite{goodfellow-et-al-2016}. They consist of two main components: an encoder function, $\mathbf{h}=f(\mathbf{x})$, which maps the input data $\mathbf{x}$ to a lower-dimensional \textit{latent representation} $\mathbf{h}$, and a decoder function, $\mathbf{r}=g(\mathbf{h})$, which reconstructs the original data from $\mathbf{h}$. A common application of autoencoders is image compression, where the encoder reduces an image to a compact latent representation, and the decoder reconstructs the image from this compressed form.

Autoencoders aim to approximately reconstruct the input data, $\mathbf{r} \approx \mathbf{x}$, as achieving exact reconstruction is often impractical. However, the strength of autoencoders lies in the compressed representation they generate, which can be leveraged for tasks such as denoising, anomaly detection, and feature extraction.

In many real-world applications, input data $\mathbf{x}$ may be corrupted by noise or other distortions, resulting in a noisy version $\mathbf{\tilde{x}}$. Despite the presence of noise, $\mathbf{\tilde{x}}$ often retains valuable information about the original data. Denoising Autoencoders \cite{vincent_dae_2008} are a specialized type of autoencoder designed to learn a robust representation of $\mathbf{x}$ from noisy input data. This is achieved by minimizing a loss function that enforces the reconstructed output to align with the desired properties of $\mathbf{x}$. In quantum mechanics, Denoising Autoencoders have been successfully applied to quantum state tomography \cite{barzili_2021,Danaci_2021}. 

To address the quantum marginal compatibility problem introduced in Section \ref{sec:compatibility_problem}, we propose a CDAE architecture. To represent density matrices in a format suitable for the CDAE, we decompose them into two-channel tensors: one channel containing the real part and the other containing the imaginary part. This representation is analogous to how images are represented as multi-channel tensors, enabling the application of computer vision techniques, such as CNNs \cite{oshea2015introduction}, to address our problem.

\begin{figure*}
    \centering
    \includegraphics[width=16cm, height=3.3cm]{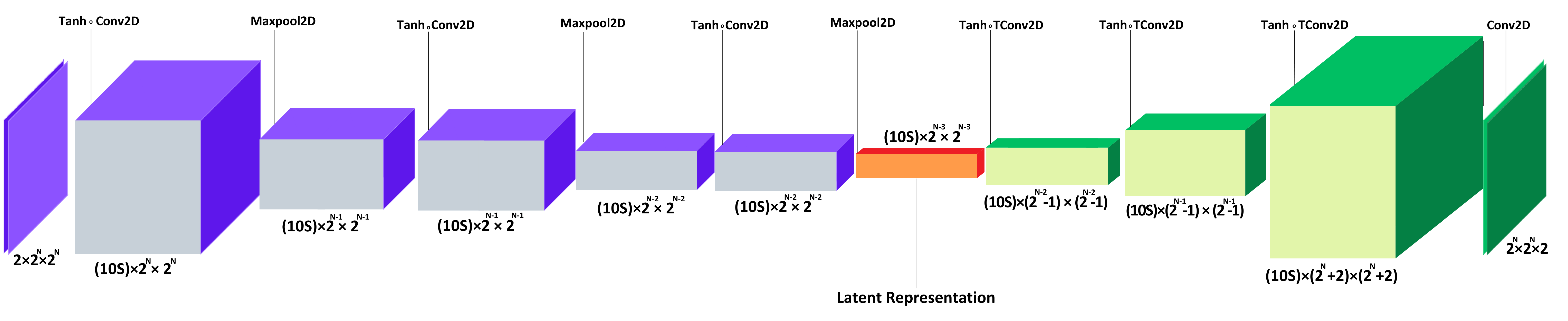}
    \caption{Illustration of the proposed CDAE architecture. The boxes represent tensors with dimensions specified as $(\textit{number of channels}) \times \textit{height} \times \textit{width}$, indicated below each box. The operation applied to the tensor on the left is written above the arrow between two tensors, with the resulting tensor shown on the right. Since the Tanh activation function does not modify tensor dimensions, we simplify the notation by combining convolutional operations ($\mathit{Conv2D}$ and $\mathit{TConv2D}$) with the $\mathit{Tanh}$ activation function into single operations, denoted as $\mathit{Tanh}\circ\mathit{Conv2D}(\ldots)$ and $\mathit{Tanh}\circ\mathit{TConv2D}(\ldots)$. All boxes and operations to the left (right) of the \emph{Latent Representation} correspond to the \emph{encoder} (\emph{decoder}). The leftmost box represents the input $\mathbf{\tilde{x}}$ of the CDAE, while the rightmost box represents its output $\mathbf{z}$.}
    \label{fig:autoencoder}
\end{figure*}

For $N$-qubit systems, each channel corresponds to a $2^N \times 2^N$ matrix. The full architectural specifications of the implemented model are detailed in \ref{app:2}. Our CDAE architecture supports any $N$-qubit system with $N > 2$, as demonstrated in \ref{app:1}. Figure \ref{fig:autoencoder} illustrates the proposed CDAE architecture. The encoder, which is responsible for extracting the latent representation, comprises the layers to the left of the latent representation in the figure. The decoder, which is tasked with reconstructing the original data from the latent code, consists of the layers to the right of the latent representation.

\section{Approach}
\label{sec:approach}

Let $\tilde{\mathbf{x}}$ and $\mathbf{z}$ be the input and output of our CDAE, respectively, such that $\mathbf{z} = g(f(\tilde{\mathbf{x}}))$. The input $\tilde{\mathbf{x}}$ is the two-channel representation of the corrupted density matrix $\tilde{X}$
\begin{equation}
    \tilde{X} = \mathcal{Q}_{\mathcal{J}_M}\circ\dots\circ\mathcal{Q}_{\mathcal{J}_1}( \rho ),
    \label{matrix_X}
\end{equation}
\noindent obtained by composing equation (\ref{mio}) for a given set $\{ \sigma_{\mathcal{J}_{\alpha}} \}_{\alpha=1}^M$ of quantum marginals, with $\rho$ serving as an initial seed state. The term 'corrupted' refers to the presence of negative eigenvalues in the matrix $\tilde{X}$, which indicates that it is not a valid quantum state, although it still contains quantum marginals, that is, $\sigma_{\mathcal{J}{\alpha}} = \mathrm{tr}_{\mathcal{J}^{c}_{\alpha}} [ \tilde{X} ]$ for all $\alpha = 1,\ldots,M$. The output $\mathbf{z}$ is also a two-channel tensor, and its matrix representation $Z$ can be obtained by taking the first channel as the real part and the second channel as the imaginary part.

In addition, we explore an approach that involves using the output of the CDAE as an initial seed for the MIO. Since the MIO is capable of reconstructing global states that match the given marginals with perfect fidelities but often suffers from the issue of producing unphysical, this hybrid strategy could mitigate this issue. By providing a well-structured initial guess from the trained autoencoder, the MIO may produce physically valid density matrices more reliably.

\subsection{Data generation}\label{data_generation}
Here, we describe the procedure used to generate the data for the training and testing stages. The set $\{ \tilde{\mathbf{x}}_{\beta}, \mathcal{M}_{\beta} \}_{\beta=1}^{N_S}$ represents our training set, with $N_S$ denoting the number of samples, and $\mathcal{M}_{\beta} = \{ \sigma_{\mathcal{J}_{\alpha}} \}_{\alpha=1}^M$. Each set $\mathcal{M}_{\beta}$ is obtained by calculating all $k$-body quantum marginals of an $N$-qubit density matrix $\rho^{(\beta)}$, where $|\mathcal{M}_{\beta}| = {N \choose k}$. We choose $k = N-1$ to generate the data in all the cases presented in this work. However, the models trained in this way also work for $0 < k<N-1$, as demonstrated in the results section. We consider $r$-rank density matrices $\rho^{(\beta)}$, generated according to the following procedure:
\begin{enumerate}    
    \item[i)] Randomly generate an integer $r$ from a uniform distribution in the interval $r_{\mathrm{min}} \leq r \leq 2^N$, where $r_{\mathrm{min}}$ is a pre-defined parameter.
    \item[ii)] Generate a probability distribution vector $\mathbf{p}_{\beta} = (p_1^{\beta}, \ldots,p_r^{\beta})^T$.
    \item[iii)] Sample $r$ pure states $\ket{\psi^{\beta}_i}$ from the Haar measure.
    \item[iv)] Form the density matrix $\rho^{(\beta)} = \sum_{i=1}^r p^{\beta}_i \kb{\psi^{\beta}_i}{\psi^{\beta}_i}$.
\end{enumerate}
Thus, we generate $\rho^{(\beta)}$ and then, using the partial trace $\sigma_{\mathcal{J}_{\alpha}} = \mathrm{tr}_{\mathcal{J}^{c}_{\alpha}} \left[ \rho^{(\beta)} \right]$, we obtain the elements of $\mathcal{M}_{\beta}$. With $\mathcal{M}_{\beta}$, we find $\tilde{X}_{\beta}$ using equation (\ref{matrix_X}) and then its two-channel representation $\tilde{\mathbf{x}}_{\beta}$. We repeat this procedure $N_S$ times to form the training set.

\subsection{The loss function}
To train our CDAE, we define a loss function that ensures that the output is a PSD matrix and compatible with the given quantum marginals. Let $Z$ be the matrix representation of the output $\mathbf{z}$ of the CDAE. We can decompose $Z$ into its polar form as $Z = UP$, where $U$ is a unitary matrix and $P$ is a PSD matrix. The loss function is given by:
\begin{equation}\label{loss}
\mathcal{L} = \ell(\mathrm{Re}(U),\mathbb{1}) + \ell(\mathrm{Im}(U),\mathbb{0}) +  \sum_i^{|\mathcal{M}_{\beta}|} \ell(\sigma_{\mathcal{J}{i}}, z_{\mathcal{J}{i}}),
\end{equation}
where $\ell(.,.)$ denotes the mean square error, $\mathbb{1}$ is the identity matrix, and $\mathbb{0}$ is the zero matrix. The first two terms in equation (\ref{loss}) ensure that the output matrix $Z$ is PSD, while the third term accounts for the quantum marginals, with $z_{\mathcal{J}{i}} = \mathrm{tr}_{\mathcal{J}^{c}_{i}} \left[ Z \right]$ representing the output marginal. Although we could explicitly include a term $\ell(\mathrm{Tr}[Z], 1)$ in equation (\ref{loss}) to guarantee normalization, we find that this constraint is approximately satisfied without the need to explicitly include it. The final output matrix can be renormalized as $Z/\mathrm{Tr}[Z]$ to ensure exact normalization. Additionally, in practice, $Z$ is approximately Hermitian, which can be fixed by setting $Z = (Z + Z^{\dagger})/2$, thus avoiding nonreal eigenvalues.

\subsection{Training}
For training, we used Adam Optimizer with a learning rate of $10^{-4}$ and a batch size of $100$ for all the cases presented. The model was implemented in PyTorch, and all numerical experiments were conducted on a workstation equipped with 125 GB of RAM, an NVIDIA GeForce RTX 4090 GPU and an AMD Ryzen Threadripper 7980X 64-core CPU. The complete code and instructions for reproduction are available in a GitHub repository \cite{repo_mlqmp}.

The architecture employed in this work demonstrates sufficient generalization capacity, successfully adapting to global quantum states ranging from 3 to 8 qubits without requiring significant structural modifications. As a result, there is no need to design new architectures or increase the model capacity for each case (e.g., by deepening the network or increasing the number of convolutional filters). This flexibility enables the use of transfer learning as a strategy for extending the model to systems with a higher number of qubits.

Transfer learning allows us to initialize training for a model targeting $N$ qubits from a previously trained model with $N-1$ qubits, rather than training from scratch. Our results indicate that this approach substantially reduces both the training time and the amount of data required. During the transfer process, certain layers, typically the encoder, are frozen, i.e., their gradients are deactivated, and only selected layers (often in the decoder) are updated. We begin with a base model trained on 3-qubit global states. Once trained to an acceptable level of performance—achieving fidelities of approximately 0.98 for a random case—we use this model as a starting point for training a model for 4-qubit global states. In this case, it was sufficient to retrain only the final decoder layer, yielding fidelities of around 0.97 for a test case. This strategy is repeated to scale up the model. However, in some cases, retraining only the decoder was insufficient. For example, when transitioning from 4 to 5 qubits, retraining the full model was required to achieve average fidelities above 0.95 for randomly selected cases. Interestingly, for larger systems (6, 7, and 8 qubits), retraining only the decoder once again proved sufficient. In these cases, transfer learning from the immediately smaller model led to successful generalization, enabling high-fidelity global state reconstruction based on marginal inputs.

These findings suggest a remarkable property of the network: the encoder appears to capture the relevant marginal information in a way that is transferable across different system sizes. This highlights the model’s ability to internalize the underlying non-linear structure of the QMP and leverage it effectively across varying dimensional scales.

\section{Results}
In this section, we present the results of our experiments to evaluate the performance of the proposed CDAE in solving the task of finding a quantum state compatible with a given set of quantum marginals. We focus on two key metrics: the success rate and accuracy. The success rate measures the fraction of outputs predicted by the model that are valid PSD matrices. 

To assess the accuracy of the predicted quantum marginals, we use the \emph{mean fidelity}, denoted $F_{\mathit{mean}}$. For each sample $\beta$ (see \ref{data_generation}) in the test set, we compute the average fidelity $\overline{F}_{\beta}$ over all the marginals in that sample as $\overline{F}_{\beta} = \sum_{i=1}^{| \mathcal{M}_{\beta}| } F( \sigma_{\mathcal{J}_{i}}, z_{\mathcal{J}_{i}})/| \mathcal{M}_{\beta}| $, with $F( \sigma_{\mathcal{J}_{i}}, z_{\mathcal{J}_{i}})= \mathrm{tr}\left[ \sqrt{\sqrt{\sigma_{\mathcal{J}_{i}}} z_{\mathcal{J}_{i}} \sqrt{\sigma_{\mathcal{J}_{i}}}} \right]^2$ the fidelity between the target marginal $\sigma_{\mathcal{J}_{i}}$ and the predicted marginal $z_{\mathcal{J}_{i}}$. Finally, the mean fidelity $F_{\mathit{mean}}$ is obtained by averaging $ \overline{F}_{\beta}$ over all samples in the test set; this is $F_{\mathit{mean}} = \sum_{\beta=1}^{N_{ \mathit{test}} } \overline{F}_{\beta}/ N_{ \mathit{test}}$, where $N_{ \mathit{test}}$ is the number of samples. To show variability in the data, we use the standard deviation (SD), calculated as $SD = \sqrt{ (N_{ \mathit{test}} - 1)^{-1} \sum_{\beta=1}^{N_{ \mathit{test}} } ( F_{\mathit{mean}} -  \overline{F}_{\beta}  )^2 }$.

We consider $N$-qubit global states with $k$-qubit marginals, denoted as $N\mathrm{\mathrm{\#}}k\mathrm {\mathrm{\#}}$. We label the model that uses the full loss function from equation (\ref{loss}) as \emph{model1}, while \emph{model2} uses equation (\ref{loss}) without the marginals term $\sum_i \ell( \sigma_{\mathcal{J}_{i}}, z_{\mathcal{J}_{i}})$. We determine the scaling parameter $S$ by experimentation, keeping the minimum value of $S$ that produces satisfactory levels of accuracy and success rate. 

\begin{figure}[htp]

\subfloat[N3k1]{%
\includegraphics[scale=0.5]{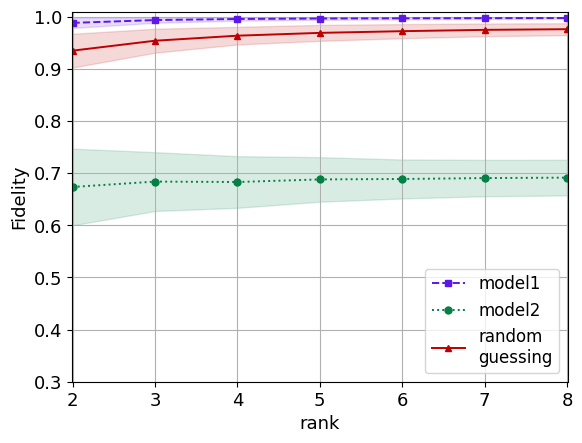}%
\label{fig:N3k1}
}
\subfloat[N3k2]{%
\includegraphics[scale=0.5]{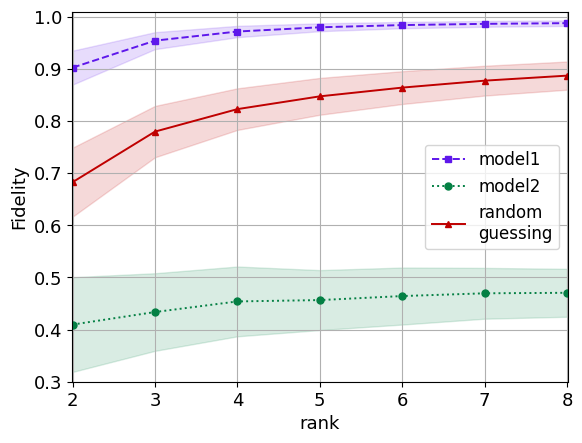}%
\label{fig:N3k2}
}
\caption{Mean and standard deviation (colored cloud) of the fidelity, obtained over $10^4$ samples, versus \emph{rank} for \emph{mode1} (purple squares), \emph{model2} (green dots) and \emph{random guessing} (red triangles) for the cases (a) $N3k1$ and (b) $N3k2$.}
\end{figure}

To evaluate the performance of \emph{model1} and \emph{model2} in the case $N3k2$, we set the scaling factor $S=10$ and trained both models for $100$ epochs using the same data set (see Section \ref{data_generation}) of size $10^6$, with weights initialized in the same seed. We then evaluated the performance of these trained models in both the $N3k1$ and $N3k2$ cases. The values of $F_{\mathit{mean}}\pm SD$, calculated over $10^4$ samples per rank, are shown in figure \ref{fig:N3k1} and figure \ref{fig:N3k2}, respectively, for ranks up to $2^N$. As expected, predictions from \emph{model1} (purple squares) consistently achieve higher fidelity with target marginals compared to \emph{model2} (green dots) for all ranks, and both models achieve a high success rate. This demonstrates the importance of including the marginals term in the loss function. Moreover, predictions from \emph{model1} outperform \emph{random guessing} (red triangles), which involve generating random $N$-qubit density matrices, calculating their $k$-qubit marginals, and computing the fidelity with the target marginals. 

\begin{figure*}[!htp]
    \centering
    \includegraphics[scale=0.455]{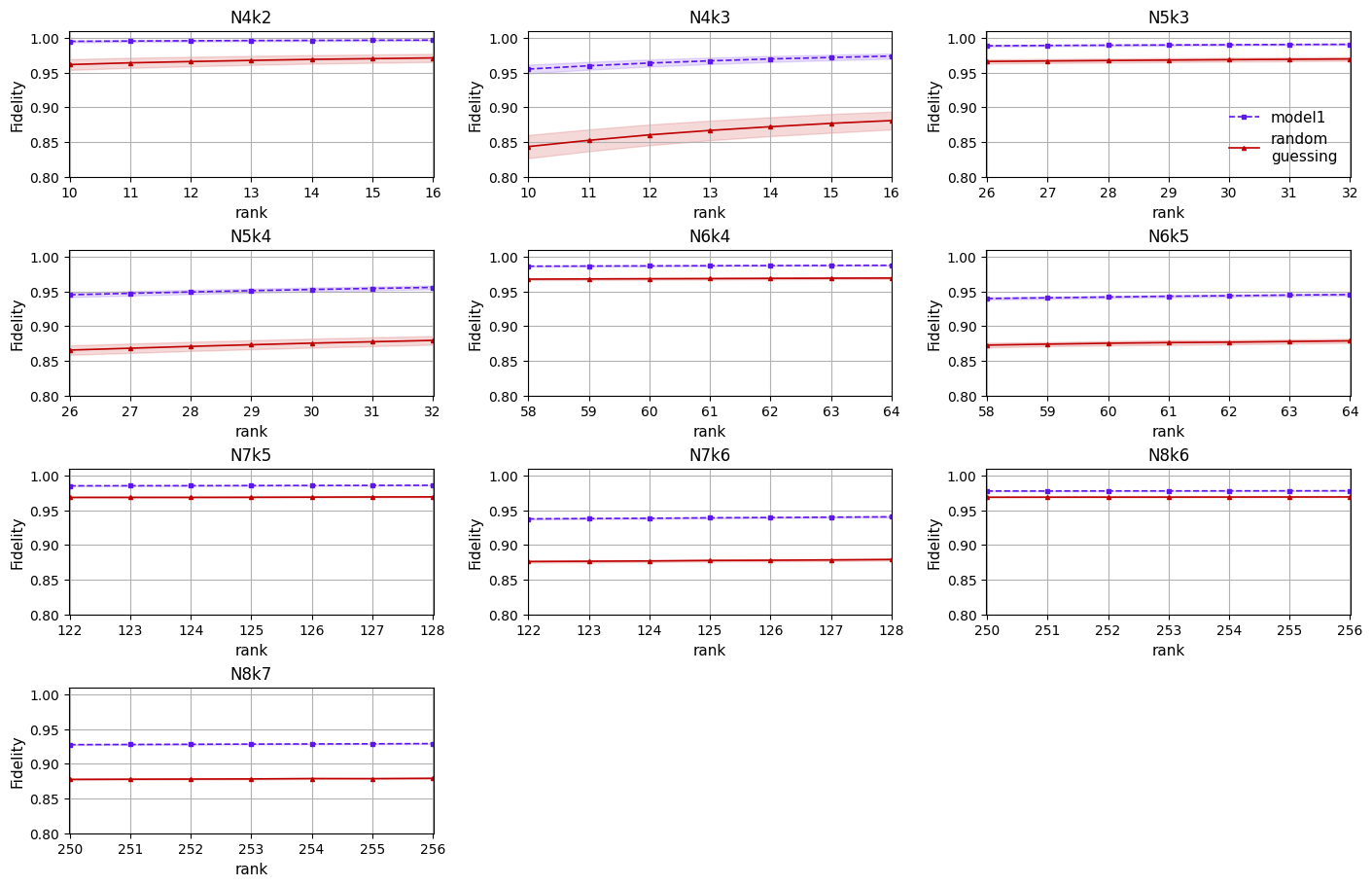}
    \vspace{-0.5cm}
    \caption{$F_{\mathit{mean}}\pm SD$, over $10^4$ samples, as a function of \emph{rank} for \emph{model1} (purple dashed line) and for \emph{random guessing} (solid red line) for the cases $N4k2$, $N4k3$, $N5k3$, $N5k4$, $N6k4$, $N6k5$, $N7k5$, $N7k6$, $N8k6$ and $N8k7$.}
    \label{fig:model-fidelities}
\end{figure*}

\begin{figure*}[!htp]
    \centering
    \includegraphics[scale=0.59]{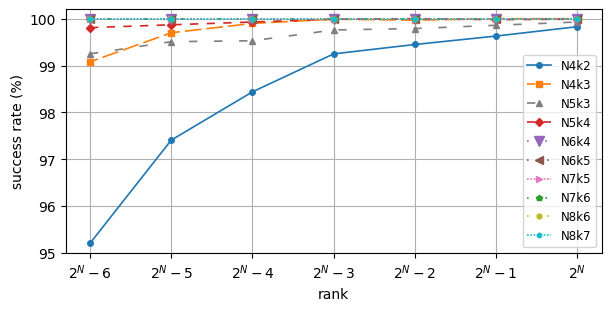}
    \vspace{-0.3cm}
    \caption{Success rates of \textit{model1} for all the cases presented in Figure~\ref{fig:model-fidelities}. \textit{model1} achieved a $100\%$ success rate for all the cases with $N>5$, with the lines overlapping at the $100\%$ mark.}
    \label{fig:succes_rates_1}
\end{figure*}

In figure \ref{fig:model-fidelities}, we compare \textit{model1} to random guessing on systems with $N>3$, up to $N=8$ qubits, using the mean fidelity $F_{\mathit{mean}}$ as a performance metric. The comparison is carried out for different values of $k$. We observe that \textit{model1} consistently outperforms random guessing in all scenarios, with the performance gap increasing as $k$ grows. This suggests that the model retains the information provided by the marginals with high accuracy. Furthermore, \textit{model1} achieved high success rates for all cases shown in figure \ref{fig:model-fidelities}: above $95\%$ for the case $N4k2$, and above $99\%$ for the remaining cases -some of which reached $100\%$- as detailed in figure \ref{fig:succes_rates_1}.

\begin{figure*}[!htp]
    \centering
    \includegraphics[width=\textwidth]{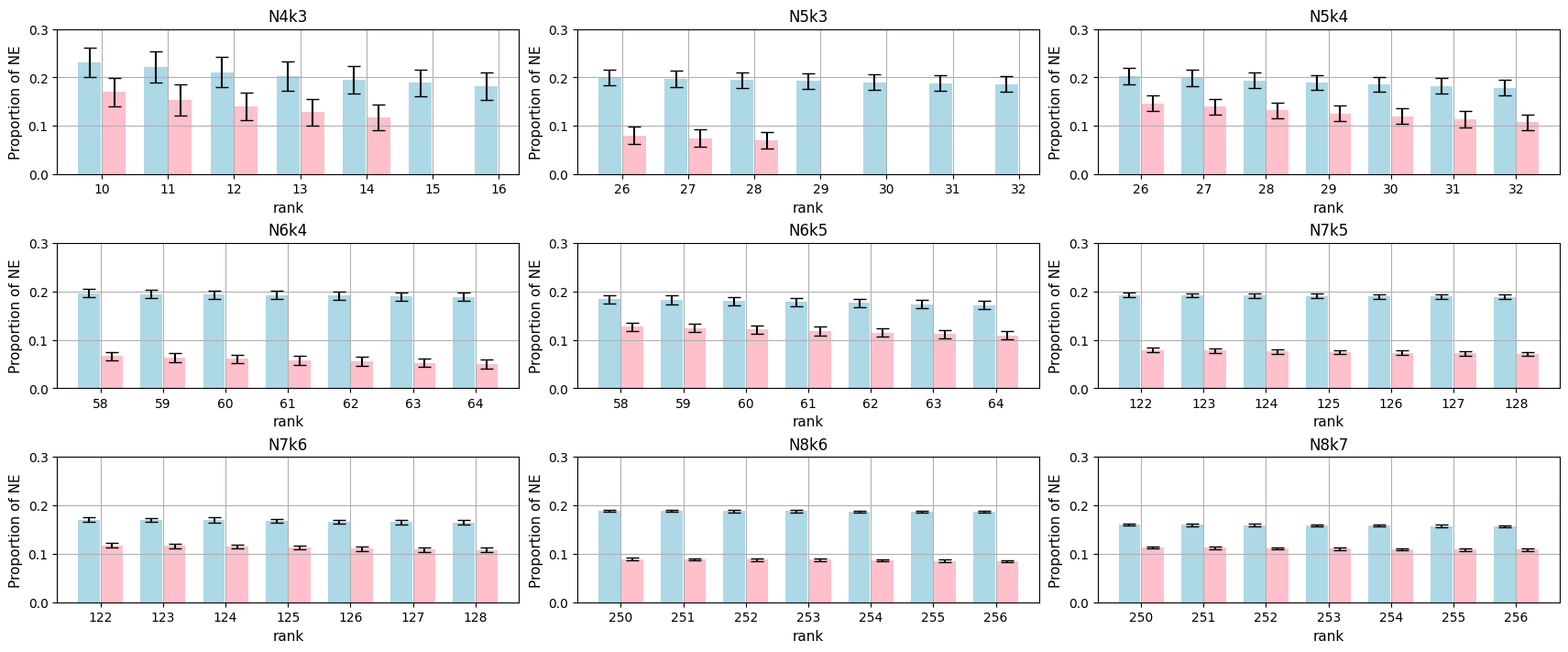}
    \vspace{-0.3cm}
    \caption{Comparison of the proportion of negative eigenvalues relative to the total number of eigenvalues. Each bar represents the average over a sample of $10^4$ tests per rank, with error bars indicating the standard deviation. Light blue bars correspond to the first pass through the MIO, while pink bars correspond to the \textit{model1}+MIO configuration.}
    \label{fig:proportion}
\end{figure*}

Moreover, we tested a scheme in which the output of \textit{model1} is passed through the MIO (\textit{model1}+MIO). By first applying the MIO and then feeding its output into the CDAE, we obtained a global state that closely reproduces the target marginals with high fidelity. This resulting state can then serve as a new seed for a subsequent pass through the MIO, as given by equation (\ref{matrix_X}), using the given target marginals $\{\sigma_{\mathcal{J}_{\alpha}} \}_{\alpha=1}^M$. Depending on the number $k$ of qubits in the quantum marginals, some corrupted (i.e. negative) eigenvalues may still persist. However, we observed that, in many cases, the final output contains no corrupted eigenvalues, or at least fewer than those present after the initial MIO pass, see Figure~\ref{fig:proportion} for details. Additionally, the magnitude of these corrupted eigenvalues tends to decrease by approximately one order of magnitude.

\begin{figure*}[!htp]
    \centering
    \includegraphics[scale=0.45]{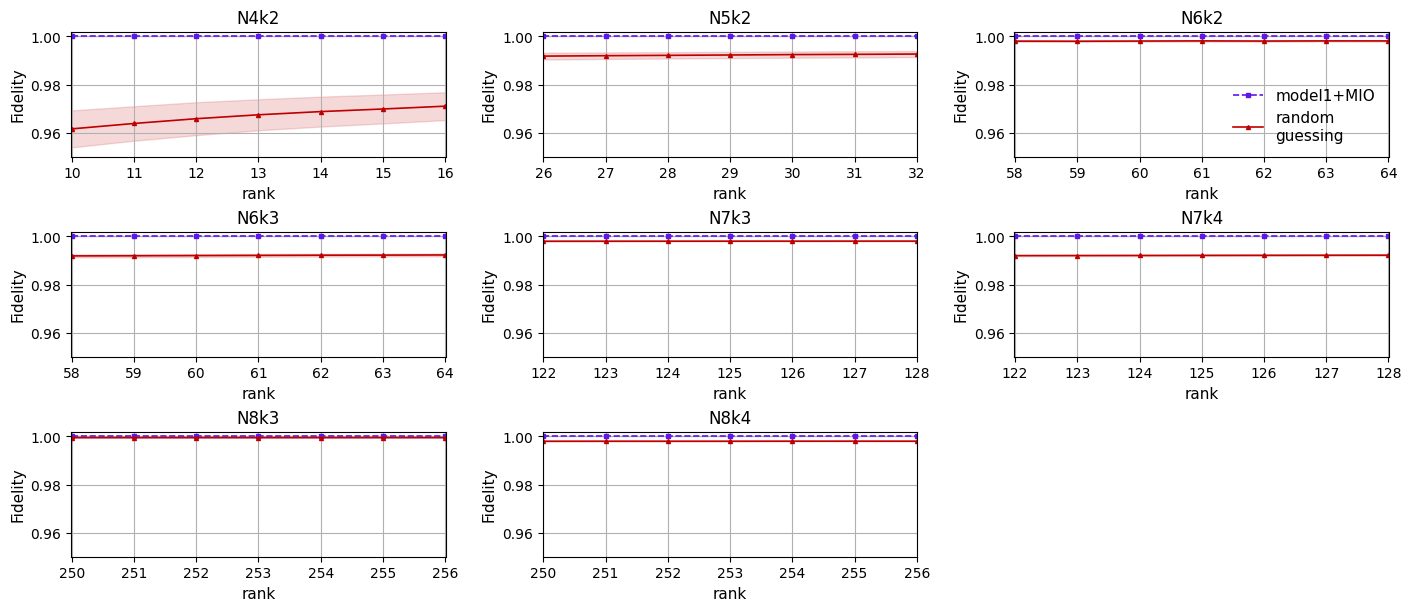}
    \vspace{-0.3cm}
    \caption{$F_{\mathit{mean}}\pm SD$, over $10^4$ samples, as a function of \emph{rank} for \emph{model1} with an additional MIO pass (purple dashed line) and for \emph{random guessing} (solid red line). Results are shown for the cases $N4k2$, $N5k2$, $N6k2$, $N6k3$, $N7k3$, $N7k4$, $N8k3$ and $N8k4$. All the states produced by the \textit{model1}+MIO scheme match the target marginals with perfect fidelity.}
    \label{fig:model-exact-fidelities}
\end{figure*}

\begin{figure*}[!htp]
    \centering
    \includegraphics[scale=0.63]{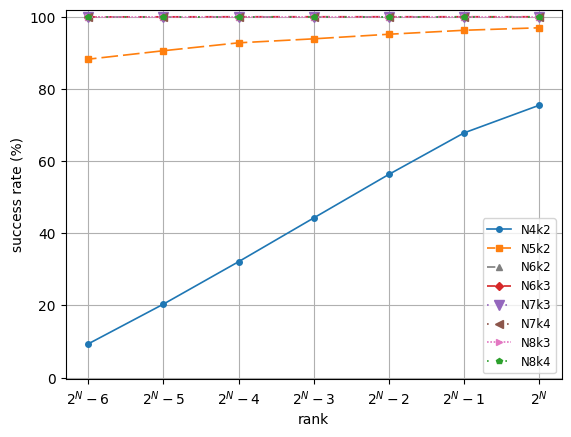}
    \vspace{-0.3cm}
    \caption{Success rates of the \textit{model1}+MIO scheme for all the cases presented in Figure~\ref{fig:model-exact-fidelities}. The scheme achieved a $100\%$ success rate for all the cases with $N>5$, with the lines overlapping at the $100\%$ mark.}
    \label{fig:succes_rates_2}
\end{figure*}

In figure \ref{fig:model-exact-fidelities}, we show results for cases in which the final output of this scheme (\textit{model1} + MIO) is a valid quantum state whose marginals match the target marginals with perfect fidelity. Success rates of these cases are shown in figure \ref{fig:succes_rates_2}. For systems with $N>5$, all the instances were perfectly reconstructed, achieving a success rate of $100\%$, whereas for the cases $N4k2$ and $N5k2$ the success rate consistenly increases with the rank. These results suggest that further improvements to the model, such as adding more layers, increasing filter sizes, or training on larger datasets, could potentially led the model to find exact solutions in all the cases. 

We experimented with iterating the CDAE and MIO steps multiple times in cases where exact solutions were not initially found, but observed no significant improvement beyond the \textit{model1}+MIO scheme. This suggests that a single iteration is sufficient to achieve this form of “purification”. We hypothesize that this is because the remaining negative eigenvalues after the second MIO pass are too small to be identified by the CDAE as noise in subsequent iterations.

\begin{figure}[htp]
    \centering
    \includegraphics[scale=0.58]{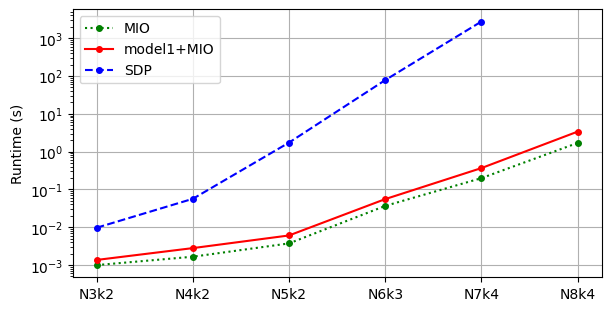}
    \vspace{-0.3cm}
    \caption{Average computation time of the CVXPY SDP solver (blue dashed line), \textit{model1} (green dotted line), and \textit{model1}+MIO (red solid line) across different system sizes. The vertical axis is in logarithmic scale. The SDP solver failed to run for the 8-qubit case.}
    \label{fig:runtime}
\end{figure}

In addition, we compare the performance of our model against state-of-the-art SDP solvers in terms of computation time. Figure \ref{fig:runtime} shows the average runtime computed over $10^3$ samples for cases with $N<6$, $100$ samples for the case with $N=6$ and $50$ samples for cases with $N>6$. In all cases, the target marginals were obtained from full-rank density matrices. As shown in figure \ref{fig:runtime}, once trained, both \textit{model1} and \textit{model1}+MIO execute faster than the SDP solver provided by the CVXPY library in \texttt{Python}. While the SDP solver consistenly found solutions matching the target marginals with perfect fidelity, it failed to run for the $N8k4$ case on our machine. In contrast, our \textit{model1}+MIO scheme successfully found solutions with exact marginals for $N8k4$, with an average runtime of approximately $3.38$ seconds.

\section{Conclusions and Discussion}
We have successfully developed a CDAE and combined it with the MIO \cite{uzcateguic_2022} to find a global $N$-qubit density matrix compatible with a given set of quantum marginals. Our model, which admits any $N$-qubit global system with more than $2$ qubits, transforms, in most cases, the output of the MIO into a PSD matrix, preserving information about the quantum marginals with high accuracy. 

Transfer learning proved to be a valuable technique in our approach, allowing us to reduce the size of the training data set and improve computational efficiency. By keeping parts of the CDAE fixed and retraining the rest of the architecture, we were able to leverage the knowledge learned in smaller-scale cases to improve performance on larger-scale problems.

Although our model demonstrates promising results, it is important to note that it may struggle with cases where the quantum marginals correspond to pure global states. Previous theoretical studies have shown that the representation of the global state is unique in such cases, making it more challenging for our model to find a specific solution. However, for mixed states, multiple representations exist, increasing the likelihood of finding an acceptable solution.

Our work makes several contributions. First, the two-channel encoding of a density matrix introduced here offers a novel approach for studying quantum states using convolutional neural networks (CNNs). Our numerical results demonstrate that CNNs are capable of capturing fundamental features of density matrices. As noted in \cite{odense2020layerwiseknowledgeextractiondeep}, the softmax layer in convolutional neural networks and autoencoders, when using tanh or ReLU activation functions, is highly explicable through rule extraction. This suggests that convolutional autoencoders, such as the one used in this work, may be useful for uncovering new insights into challenging problems such as the QMP. Moreover, autoencoders have been successfully applied as anomaly detectors in various domains \cite{aes_anomaly_detectio_cloud_2020}. Future work could explore the use of convolutional autoencoders to distinguish between compatible quantum marginals and anomalies (incompatible marginals).

We demonstrated that combining the MIO with a CDAE enables the construction of quantum states compatible with a given set of quantum marginals, preserving them with high accuracy. In specific cases, a second application of the MIO further improves the result, yielding quantum states whose marginals match the inputs exactly. Although training the CDAE can be computationally intensive, once trained, the inference process is significantly faster compared to current state-of-the-art SDP solvers, such as those provided by the CVXPY Python library. Therefore, in the long term, our approach presents a promising and more efficient alternative to existing SDP-based methods.

The ability of our model to generate accurate approximations might serve to provide \emph{initial guess} to solve the SDP of equation (\ref{SDP-primal}), potentially reducing its computational cost and accelerating its convergence. The technique of using an initial guess close to the solution is called \emph{warm-start} and is an active subject of research. Although warm-start  has found difficulties in being implemented in SDPs, recent work \cite{angell2024fastscalablewarmstartsemidefinite} shows its viability. 

\section*{Acknowledgments}
This work was supported by ANID grants No.\thinspace$2023-3230427$ and No.\thinspace$2022-21221096$, and the ANID Millennium Science Initiative Program $\mathrm{ICN17}\_\mathrm{012}$. We also thank Proyecto FONDECyT Regular No.\thinspace1230586, and Dr.\thinspace Dardo Goyeneche for his feedback.

\appendix
\section{Model architecture}
\label{app:2}
In this section, we describe the specifics of the proposed CDAE. The encoder consists of multiple layers, incorporating 2D convolution (\emph{Conv2D}) operations, 2D max pooling (\emph{MaxPool2D}) operations \cite{dumoulin2018guideconvolutionarithmeticdeep}, and Hyperbolic Tangent (\emph{Tanh}) activation functions. The decoder includes \emph{Conv2D} operations and \emph{Tanh} activation functions, along with layers utilizing 2D transposed convolutions (\emph{TConv2D}).

The tables \ref{tab:encoder} and \ref{tab:decoder} provide detailed specifications of each layer in the encoder and decoder, respectively. All convolutional and transposed convolutional layers use a dilation factor of one. In the decoder, the output padding for transposed convolutions is fixed at zero. The intermediate number of channels in both encoder and decoder is controlled by a scaling factor $S$, which allows for flexible adjustment of the model’s capacity. Lower values of $S$ reduce the number of trainable parameters, offering a trade-off between model complexity and computational efficiency.
\begin{table}[ht]
  \caption{\label{tab:encoder}Specifications for each encoder layer. Dilation is fixed to one throughout the network.}
  \begin{indented}
    \item[]\begin{tabular}{@{}ccccccc}
      \br
      Layer & Operation & Input channels & Output channels & Kernel size  & Stride & Padding \\
      \mr
      (0)   & Conv2D    & 2              & $S \times 10$   & $3 \times 3$ & 1      & 1       \\
      (1)   & Tanh      & --             & --              & --           & --     & --      \\
      (2)   & MaxPool2D & --             & --              & $2 \times 2$ & 2      & 0       \\
      (3)   & Conv2D    & $S \times 10$  & $S \times 10$   & $3 \times 3$ & 1      & 1       \\
      (4)   & Tanh      & --             & --              & --           & --     & --      \\
      (5)   & MaxPool2D & --             & --              & $2 \times 2$ & 2      & 0       \\
      (6)   & Conv2D    & $S \times 10$  & $S \times 10$   & $3 \times 3$ & 1      & 1       \\
      (7)   & Tanh      & --             & --              & --           & --     & --      \\
      (8)   & MaxPool2D & --             & --              & $2 \times 2$ & 2      & 0       \\
      \br
    \end{tabular}
  \end{indented}
\end{table}
\begin{table}[ht]
  \caption{\label{tab:decoder}Specifications for each decoder layer. Dilation is set to one, and output padding in transposed convolutions (\emph{TConv2D}) is fixed at zero.}
  \begin{indented}
    \item[]\begin{tabular}{@{}ccccccc}
      \br
      Layer & Operation & Input channels & Output channels & Kernel size  & Stride & Padding \\
      \mr
      (0)   & TConv2D   & $S \times 10$  & $S \times 10$   & $3 \times 3$ & 2      & 1       \\
      (1)   & Tanh      & --             & --              & --           & --     & --      \\
      (2)   & TConv2D   & $S \times 10$  & $S \times 10$   & $5 \times 5$ & 2      & 1       \\
      (3)   & Tanh      & --             & --              & --           & --     & --      \\
      (4)   & TConv2D   & $S \times 10$  & $S \times 10$   & $6 \times 6$ & 2      & 0       \\
      (5)   & Tanh      & --             & --              & --           & --     & --      \\
      (6)   & Conv2D    & $S \times 10$  & 2               & $3 \times 3$ & 1      & 0       \\
      \br
    \end{tabular}
  \end{indented}
\end{table}

\section{The CDAE compatibility}
\label{app:1}
This appendix provides a detailed analysis of the channel sizes in the CDAE presented in section \ref{sec:approach}. Each input to the CDAE is a two-tensor channel, where each channel is a $2^N\times2^N$ matrix. We demonstrate that for $N>2$, the channel size of the CDAE's output remains consistent at $2^N\times2^N$.
The size of a channel can change when \emph{Conv2D} operations, \emph{MaxPool2D} operations and \emph{TConv2D} operations are applied to tensors at different layers of the CDAE. Our model considers a setting with square channels, square kernels, the same stride along both axes and the same padding along both axes. Under these conditions, the channel size for \emph{Conv2D} and \emph{MaxPool2D} operations can be calculated using the following formula \cite{dumoulin2018guideconvolutionarithmeticdeep}
\begin{equation}
  \mathtt{C}^{E/D}_{out(j)} =  \floor{\frac{ \mathtt{C}^{E/D}_{in(j)} + 2 \mathcal{P} - \mathcal{D} (\mathcal{K} - 1) -1}{\mathcal{S}} + 1 },
  \label{layer_output_size}
\end{equation}
with $\lfloor \ldots \rfloor$ the floor function. $\mathtt{C}^{E/D}_{in(j)}$ and $\mathtt{C}^{E/D}_{out(j)}$ are the input and output size, respectively, of a channel at layer $(j)$ of the encoder($E$)/decoder($D$). $\mathcal{P}$ is the padding, $\mathcal{D}$ the dilation, $\mathcal{K}$ the kernel size and $\mathcal{S}$ the stride. The dilation is set to $\mathcal{D}=1$ across the entire CDAE.

Note that $\mathtt{C}^{E/D}_{in(l)} = \mathtt{C}^{E/D}_{out(l-1)}$ for $l>0$. This is, the input size of layer $(l)$ is equal to the output size of the previous layer $(l-1)$. The input size of the encoder is $\mathtt{C}^{E}_{in(0)}=2^N$. Let us now determine its output size $\mathtt{C}^{E}_{out(8)}$ using the values for $\mathcal{P}$, $\mathcal{D}$, $\mathcal{K}$ and $\mathcal{S}$  given in Table~\ref{tab:encoder}.
\begin{eqnarray}
  \mathtt{C}^{E}_{out(0)} &=  \Big\lfloor \frac{\mathtt{C}^{E}_{in(0)} + 2 \times 1 - 1 \times (3 - 1) -1 }{1} + 1 \Big\rfloor, \nonumber \\
  \mathtt{C}^{E}_{out(0)} &= \Big\lfloor \mathtt{C}^{E}_{in(0)} \Big\rfloor = 2^N.
\end{eqnarray}
The size of the channels is not affected by \emph{Tanh} operations. Thus, $\mathtt{C}^{E}_{out(1)} = \mathtt{C}^{E}_{out(0)}$ and therefore
\begin{eqnarray}
  \mathtt{C}^{E}_{out(2)} &=  \Big\lfloor \frac{ \mathtt{C}^{E}_{out(1)}  + 2 \times 0 - 1 \times (2 - 1) -1 }{2} + 1 \Big\rfloor, \\
  \mathtt{C}^{E}_{out(2)} &= \Big\lfloor 2^{-1}\mathtt{C}^{E}_{out(1)} - 1 + 1 \Big\rfloor = \Big\lfloor2^{-1}\mathtt{C}^{E}_{out(1)} \Big\rfloor ,\\
  \mathtt{C}^{E}_{out(2)} &= 2^{N-1}. \nonumber
\end{eqnarray}
Then, we calculate $\mathtt{C}^{E}_{out(3)}$, which is similar to $\mathtt{C}^{E}_{out(0)}$
\begin{equation}
  \mathtt{C}^{E}_{out(3)} = \floor{\mathtt{C}^{E}_{out(2)}} = 2^{N-1}. \nonumber
\end{equation}
Now, using $\mathtt{C}^{E}_{out(4)} = \mathtt{C}^{E}_{out(3)}$, we calculate $\mathtt{C}^{E}_{out(5)}$
\begin{equation}
  \mathtt{C}^{E}_{out(5)} = \floor{2^{-1}\mathtt{C}^{E}_{out(4)}} = 2^{N-2}, \nonumber
\end{equation}
and then $\mathtt{C}^{E}_{out(6)}$
\begin{equation}
  \mathtt{C}^{E}_{out(6)} = \floor{\mathtt{C}^{E}_{out(5)}} = 2^{N-2}. \nonumber
\end{equation}
Finally, the output size of the encoder $\mathtt{C}^{E}_{out(8)}$ is
\begin{equation}
  \mathtt{C}^{E}_{out(8)} = \floor{2^{-1}\mathtt{C}^{E}_{out(7)}} = 2^{N-3}. \nonumber
\end{equation}
with $\mathtt{C}^{E}_{out(7)} = \mathtt{C}^{E}_{out(6)}$. Note that for $N<3$ the encoder would render an invalid size.
Now, let us calculate the channel's output size of the decoder using the values given in Table~\ref{tab:decoder}. This is, the output size at the layer $(6)$ of the decoder. For \emph{TConv2D}, the channel's size $\mathtt{D}_{out(j)}$ changes according to \cite{dumoulin2018guideconvolutionarithmeticdeep}
\begin{equation}
  \mathtt{D}_{out(j)} = \left( \mathtt{D}_{in(j)} - 1 \right) \mathcal{S} - 2\mathcal{P} + \mathcal{D}(\mathcal{K}-1) + \mathcal{P}_{out} + 1,
  \label{tconv_output_size}
\end{equation}
where $\mathcal{P}_{out}$ is the output padding, which is set to zero across the entire CDAE. As for the encoder, $\mathtt{D}_{in(l)} = \mathtt{D}_{out(l-1)}$ for $l>0$. The input size for the first layer of the decoder is $\mathtt{C}^{E}_{out(8)}$. Thus,
\begin{eqnarray}
  \mathtt{D}_{out(0)} &= \left( \mathtt{C}^{E}_{out(8)} - 1 \right) 2 - 2 + 1(3-1)  + 1, \\
  \mathtt{D}_{out(0)} &= \left( 2^{N-3} - 1 \right) 2 + 1, \\
  \mathtt{D}_{out(0)} &= 2^{N-2} - 1. \nonumber
\end{eqnarray}
Now, knowing that $\mathtt{D}_{out(1)} = \mathtt{D}_{out(0)}$, let us determine $\mathtt{D}_{out(2)}$
\begin{eqnarray}
  \mathtt{D}_{out(2)} &= \left( \mathtt{D}_{out(1)} - 1 \right) 2 - 2 + 1(5-1) + 1, \\
  \mathtt{D}_{out(2)} &= \left( 2^{N-2} - 1 - 1 \right) 2 - 2 + 4 + 1, \\
  \mathtt{D}_{out(2)} &= 2^{N-1} -1, \nonumber
\end{eqnarray}
and then,
\begin{eqnarray}
  \mathtt{D}_{out(4)} &= \left( \mathtt{D}_{out(3)} - 1 \right) 2 - 2\times 0 + 1(6-1) + 1, \\
  \mathtt{D}_{out(4)} &= \left( 2^{N-1} - 1 - 1 \right) 2 + 5 + 1, \\
  \mathtt{D}_{out(4)} &= 2^{N} + 2, \nonumber
\end{eqnarray}
where $\mathtt{D}_{out(3)} = \mathtt{D}_{out(2)}$. Finally, we calculate the output size at layer $(6)$ using (\ref{layer_output_size})
\begin{eqnarray}
  \mathtt{C}^{D}_{out(6)} &=  \floor{\frac{\mathtt{D}_{out(5)} + 2 \times 0 - 1 (3 - 1) -1}{1} + 1 }, \\
  \mathtt{C}^{D}_{out(6)} &= \floor{\mathtt{D}_{out(5)} - 2 }, \\
  \mathtt{C}^{D}_{out(6)} &=  \floor{2^{N} + 2 - 2 } =  2^{N}, \nonumber
\end{eqnarray}
where $\mathtt{D}_{out(5)} = \mathtt{D}_{out(4)}$. Thus, we have shown that the CDAE input and output sizes are equal.

\section*{References}
\bibliographystyle{unsrt}
\bibliography{bio}
\end{document}